\documentclass[10pt,conference]{IEEEtran}
\IEEEoverridecommandlockouts
\usepackage{cite}
\usepackage{amsmath,amssymb,amsfonts}
\usepackage{algorithmic}
\usepackage{graphicx}
\usepackage{textcomp}
\usepackage{xcolor}
\usepackage{caption}
\usepackage{tabu}
\usepackage{svg}
\usepackage{url}
\usepackage{tabularx}
\usepackage{blindtext}
\usepackage{subcaption}
\usepackage{longtable}
\usepackage{array} 
\usepackage{hyperref}
\usepackage{booktabs}

\usepackage{color, colortbl}
\usepackage{adjustbox}
\definecolor{Gray}{gray}{0.9}

\def\BibTeX{{\rm B\kern-.05em{\sc i\kern-.025em b}\kern-.08em
    T\kern-.1667em\lower.7ex\hbox{E}\kern-.125emX}}
\begin{document}

\newcommand{\SMim}[1]{\textcolor{red}{{\bfseries [[#1]]}}}
\newcommand{\BJohnson}[1]{\textcolor{purple}{{\bfseries [[#1]]}}}
\newcommand{\JSmith}[1]{\textcolor{blue}{{\bfseries [[#1]]}}}
\newcommand{\fv}[1]{\textcolor{green}{{\bfseries [[#1]]}}}

\title{What Makes a Fairness Tool Project Sustainable in Open Source?}
\author{
    \IEEEauthorblockN{Sadia Afrin Mim\IEEEauthorrefmark{1}, Fatemeh Vares\IEEEauthorrefmark{2}
     Andrew Meenly\IEEEauthorrefmark{3}, Brittany Johnson \IEEEauthorrefmark{4}\\
    \IEEEauthorblockA{\IEEEauthorrefmark{1,}\IEEEauthorrefmark{2,}\IEEEauthorrefmark{4}Department of Computer Science, George Mason University}
    \IEEEauthorblockA{\IEEEauthorrefmark{2}Department of Computer Science, Rochester Institute of Technology}
    \IEEEauthorblockA{
    \IEEEauthorrefmark{1}safrinmi@gmu.edu,
    \IEEEauthorrefmark{2}fvares@gmu.edu,
    \IEEEauthorrefmark{3}axmvse@rit.edu,
    \IEEEauthorrefmark{4}johnsonb@gmu.edu
    }
    }
    }

\maketitle

\begin{abstract}

As our society grows increasingly reliant on artificial intelligence, the need to mitigate risk and harm has become paramount. 
In response, researchers and practitioners have rallied to provide tools that help mitigate risks and harms by detecting and mitigating undesired bias, commonly known as fairness tools.
Many of these tools are made available free to the public for use and adaptation.
The increase in available tooling is promising, however, little is known about the landscape of tools available beyond the ones commonly discussed (e.g., AI Fairness 360, Fairlearn). Since fairness is an ongoing necessity, the tools must be built for long-term sustainability.
Using an existing set of fairness tools as a guide, we systematically searched for the current state of art in GitHub and identified 50 projects.
We then analyzed various aspects of these tools' repositories to assess the community engagement (are developers using or engaging with them) and maintenance of these tools in practice.
Our findings suggest diverse kinds of engagements with these tool repositories, indicating their embrace of open source. Furthermore, we found that some tools are more well-maintained than others, which may affect usage in practice. Our findings revealed that 53\% of fairness projects become inactive within the first three years. By examining project sustainability in fairness tooling, we hope to bring more health and stability to growing this field.
\end{abstract}

 
 \section{Introduction}

Modern society continues to rely more and more on artificial intelligence (AI) and machine learning (ML) software and technologies as solutions to improve our day to day lives.
By some measures, ML applications already outperform humans in safety-critical domains~\cite{iqbal2022assurance} such as banking, and healthcare.
The promises of intelligent technologies continue to change the way we think about software products, applications for technology, as well as the way we build them~\cite{bini2018artificial, shah2019towards, wan2019does}.

While the potential for AI and ML are evident, we have seen numerous examples over the years of these solutions perpetuated undesired bias: from judging creditworthiness~\cite{lipsitz2006possessive} and beauty~\cite{mehrabi2021survey} to discrepancies in employee recruitment~\cite{fitria2021gender} and natural langauge translation~\cite{fitria2021gender, hasan2024olf}. 
As a result, researchers and practitioners have developed tools to help detect and mitigate these biases called fairness tools~\cite{brun2018software, lee2021landscape}. 
Fairness tools leverage a range of fairness metrics and algorithms to proactively support addressing undesired  model bias. 
Fairness tools detect bias in demographics, promote fair alternatives to existing algorithms, and encourage transparency and accountability in model decision-making processes.

A biased model can be devastating to people who are oppressed, in the minority, or otherwise vulnerable.~\cite{lee2021landscape}. As the landscape of available tools continues to expand, the challenge of locating and comparing these tools to identify the most suitable ones becomes increasingly complex.
Mim et al. have begun this work by classifying and structuring a range of available fairness tools to help users understand their scope and application~\cite{mim2023taxonomy}.
Understanding community engagement with fairness tools is crucial for identifying practical challenges, guiding improvements, and ensuring tools remain reliable and up-to-date. Insights on community engagement can also help create a healthier, stronger ecosystem for fairness tools that can support the identification of active projects, prioritize maintenance, and focus on areas needing improvements to increase relevance and usability.
In the current state of the art, we investigated the presence and use of machine learning fairness tools in open source.

\textit{The goal of this work is to understand and support the sustainability of open source fairness tool projects.}
To accomplish this goal, we conducted a mining study to extend and analyze a dataset of open source fairness tools.
Following the curation of 61 publicly available open source fairness tools, we analyzed each repository to measure engagement from current and potential contributors, as well as the maintenance efforts that may be influencing this engagement. 
We leveraged existing metrics and models to make determinations regarding project status, contributions, and other relevant activity (e.g., social engagement). 
We found that while there is a diversity of fairness tooling available in open source, only a small subset has seen meaningful or consistent engagement
We observed the highest and most consistent engagement and maintenance activity in projects affiliated with industry or non-academic organizations.
However, our findings suggest other factors may play a role in engagement and sustainability of solutions such as providing solutions for niche or popular domains (e.g., healthcare or natural language processing).
\footnote{A revised version of this paper has been accepted to the ResponsibleSE track at FSE 2025.}

The contributions of this paper are:
\begin{itemize}
    \item A publicly available dataset that catalogs 61 currently available open source fairness projects from industry, academia, and individual contributors.
    \item An empirical evaluation of engagement with and maintenance of fairness projects in open source that reveals considerations relevant to sustainable engagement. 
\end{itemize}

This paper is organized as follows: first, we review the related literature in Section \ref{related} and explain our methods for analyzing the repositories in Section \ref{method}. We then discuss the findings of our analysis in Section \ref{result} and implications from our efforts in Section \ref{discussion}. 
We finish with some limitations of our work in Section \ref{validity} and finally conclude the paper in Section \ref{conclude}.

\section{Related Work}
\label{related} 
As our work is focused on fairness tools in open source specifically on GitHub, we discuss prior work related to GitHub repository mining, ethics in open source, and fairness tool evaluations.

\subsection{Measuring Open Source Engagement and Maintenance}

Researchers have conducted a variety of exploratory studies that leverage interactions and artifacts in GitHub repositories. 
Most relevant to our efforts are prior works that mine GitHub repositories to better understand how we can measure and assess maintenance and engagement of open source repositories.
Dabic et al. explored 735,669 repositories containing 25 features to find commonly used project selection criteria for MSR studies\cite{dabic2021sampling}. 
Yang et al. in their work\cite{yang2023users} curated 576 repositories from the PapersWithCode platform to better understand the common issues reported and provided suggestions for developers to improve the overall quality of open source AI repositories and the way issues are managed inside them.
Ait et al. analyzed collected 1,127 GitHub repositories from four different classes over a period of 6 years to document behaviors over time and monitor changes over repository lifetime from four frameworks (NPM, R, WordPress and Laravel)~\cite{ait2022empirical}. 
Eluri et al. investigated factors that correlate with new contributors to open source repositories becoming long-time contributors based on the repository and contributor meta-data collected from GitHub~\cite{eluri2021predicting}. 
Coelho et al. leveraged existing repositories to propose metrics and models for measuring and predicticing the maintenance level of GitHub projects using hyperparameters(e.g, commits, pull-requests, forks) using top 10,000 starred repositories. \cite{coelho2020github}.
Jarczyk et al. analyzed GitHub repositories and proposed matrices to define the quality of open source software by measuring engagement from high volume Internet users.~\cite{jarczyk2014github}.

Hata and their colleagues investigated meta-maintenance of repository files and how similar files across repositories evolve over time using repositories of seven different programming languages~\cite{hata2021same}. 
Businge et al. investigated repository maintenance in fork-based software families within open source software ecosystems centered on Android, .NET, and JavaScript~\cite{businge2022reuse}. 
Savarimuthu et al.\cite{savarimuthu2014towards} discussed findings from a case study on norm compliance in three Java open source projects and introduced an architecture for mining norms from such projects. 

Robles et al. introduced a new method based on repository data and a survey of open source deveopers for estimating software development effort using data from source code management repositories, applied to the OpenStack project~\cite{robles2014estimating}. 
Yu et al. analyzed factors for measuring maintainability in open source, focusing on issue tracking systems, change logs, and source code~\cite{yu2005measuring} for open source maintenance from Bugzilla. These studies collectively highlight the significance of mining GitHub repositories to assess maintainability and community engagement, showcasing diverse methodologies across various open source ecosystems. They reveal key factors, such as contributor activity, issue management, and code evolution, as critical indicators for sustaining open source projects effectively over time.

 
\subsection{Ethics in Open Source}
While there may not be an abundance of literature at the intersection of open source and fairness, there have been prior efforts aimed at understanding ethical considerations in open source. 
Widder et al. conducted an empirical investigation into the how open source contributors think about ethics and the considerations made in an open source AI-based project~\cite{widder2022limits}.
Fitzgerald et al. administered a survey of Artificial General Intelligence Projects for Ethics, Risk, and Policy in open source based on seven attributes~\cite{fitzgerald20202020}. 
Franse et al. discussed operational, ethical, legal, and governance challenges associated with the application of Machine Learning (ML) have prompted a requirement for a well-defined and considerate collection of guidelines for the responsible governance, management, and implementation of ``responsible ML''~\cite{franse2022practical}. While examining ethical concerns in open source has been published in the literature, we see a gap between the literature and the language of modern open source tools. Specifically, the tools of this study use the term \textit{fairness} whereas the literature does not largely address this term.

\subsection{Fairness Tool Evaluations}

Researchers have conducted several studies to evaluate existing fairness tool support.
Mehrabi et al. investigated real-world examples of machine learning bias and established a taxonomy for defining fairness in different contexts~\cite{mehrabi2021survey}. 
Richardson et al. interviewed 20 ML practitioners to elicit their perceptions of existing fairness tools, as well as recommendations for improving tool support~\cite{richardson2021towards}.
Lee et al. built a taxonomy of machine learning fairness tools based on a mixed methods study with industry practitioners that illuminated gaps in existing tools that impact their ability to meet practitioner needs~\cite{lee2021landscape}.
Building on prior work, Mim et al. developed a taxonomy of machine learning fairness tools based on features and support across the various machine learning phases~\cite{mim2023taxonomy}.  \\

While many studies have focused on open source analysis, ethics, or mining practices within repositories, we have no knowledge of any prior research examining the landscape of open source fairness tool support in terms of their varied use and upkeep.

\section{Methodology}
\label{method}
The goal of this work is to enhance the sustainability of open source fairness tool projects by analyzing their history of maintenance practices and community involvement.
More specifically, this project aimed to assess engagement with these tools and use of these tools in practice.
Therefore, our research questions are:

\begin{description}
   
    \item[\textbf{RQ0}] \textit{What fairness tools have been released as open source projects?}
    \item[\textbf{RQ1}] \textit{To what extent have open source fairness tool projects engaged with their community?}
    \item[\textbf{RQ2}] \textit{ How active is the maintenance of open source fairness tool projects?}
    \item [\textbf{RQ3}]\textit{How long do open source fairness tool projects stay active?}
\end{description}

To answer these research questions, we began with a previously curated dataset of fairness tools~\cite{mim2023taxonomy}, which we call \textit{Original Toolset}. From the set of tools we curated, we created a list of keywords related to fairness. Using these keywords, we leveraged the GitHub API to identify and analyze additional repositories containing these terms. This process yielded a new set of tools, which we referred to as the \textit{Emergent Toolset}.





\subsection{Fairness Tool Keyword Extraction}

To drive our keyword extraction, we focused on two kinds of keywords: 1) natural language terms that relate to fairness and 2) technical keywords that related to tool specification and use.
We collected the natural language keywords from the \textbf{About} section of the GitHub repository of each fairness tool in the \textit{Original Toolset}. 
We focused on keywords that relate to fairness, such as ethics, responsibility, transparency, bias, and trust. 
We extracted tool-specific keywords from each tool's documentation, repositories, and online resources.
We examined import statements and tool component names (e.g, \textit{import fairlearn}, \textit{import aif360}).
These keywords helped us identify the usages of each tool as well as identify other relevant open source tooling.
In our keyword list, we focused on \textit{import} statements found in the tool documentation to increase the likelihood of finding if and how these tools are being used. 
We also included statements such as ``\texttt{from x import y}'' which represent alternative ways developers may be integrating these tools. For keyword analysis we only considered Python-based tools as  from our observation, most of the fairness tools are developed in Python.
We finalized our list of keywords by determining common keywords across both natural language and tool-specific keywords, which is shown in Table-\ref{keywords}.

\begin{table}[h]
\centering
\caption{Keywords Extracted from \textit{Original Toolset}}
\begin{tabular}{l l}
\toprule
\textbf{Keyword Type} & \textbf{Keywords Curated} \\ 
\midrule
\rowcolor{Gray}
\textit{Natural Language} & \begin{tabular}{@{}l@{}}Ethics,ML fairness, ML Ethics, Ethical ML,\\
Machine Learning Bias, AI bias, AI Fairness,\\ Fairness metrics, Bias mitigation, Model Fairness,\\ Model Bias, Fairness Evaluation, Biased Dataset, \\Fairness Algorithm, Fairness toolkit, \\AI Transparency tool, ML Transparency tool \end{tabular} \\ 

\textit{Tool Specification}  & \begin{tabular}{@{}l@{}} TensorFlow fairness,
IBM fairness,\\ Microsoft fairness, import aequitas, aequitas,\\ import themis-ml, themis-ml,drivendataorg/deon,\\ fairkit-learn, fat-forensics,\\ import fairness indicator, fairness indicator,\\ import WitWidget, WitConfigBuilder,
import aif360, \\aif360, fairlearn,import fairlearn, AI Fairness 360,\\ What-if tool google, Linkedin LiFT \end{tabular}\\ 
\bottomrule
\end{tabular}
\label{keywords}
\end{table}

\subsection{Manual Keyword Validation}
To validate our set of keywords and its ability to return the desired results, we first conducted a manual evaluation of our search results.
We examined the variety of artifacts found, including repositories, issues, and commits to see if they would return additional tools from the \textit{Original Toolset}.
Given the promising results from our manual analysis, and the time-consuming nature of this approach, we wrote a script to automate the rest of our search.


\subsection{Automated Data Collection}
Given the outcomes of our manual validation, we created automated scripts to support the curation and storage of our search results. 
We quickly realized how large the volume of the search results (ranging from 300 to 13,200) for any given keyword, let alone the aggregation of them all.
Therefore, to collect results consistently across repositories, we set of threshold of top-10 search results for each keyword as only the top 7-8 of the returned valid tools that explicitly works on Machine Learning fairness.
We wrote Python scripts that use the Github API\footnote{\url{https://docs.github.com/en/rest?apiVersion=2022-11-28}} to return keyword search results. 
Given the Github API provides access to code, repositories, issues, commits, and pull requests, we focused our search and analyses on these artifacts.

\subsection{Augmenting \textit{Original Toolset(RQ0)}}
Our automated search returned a diverse set of repositories and contexts in which our keywords appeared. We found three types of results from the \textit{Original Toolset}: \textit{new tools/libraries/algorithms}, \textit{demos/tutorials}, and \textit{third-party integration}.
While our keywords often returned artifacts relevant to our \textit{Original Toolset}, we also found that our keywords uncovered additional fairness tools not in our original set. We will refer to the new set of tools that we found in this analysis as the \textit{Emergent Toolset}. 

\subsection{Evaluating Community Engagement (RQ1)}\label{eval_engagement}
To investigate community engagement, we examined the following metrics:
\begin{itemize}
    \item Development Source
    \item Availability as a Research Article
    \item Trends of Watches, Forks, Stars and Pull requests
\end{itemize}
We manually classified the origin of the repositories as originating from academic research, industrial research or individual effort examining the affiliation of the developers of the project and repository owner description. We determined their availability as research articles by examining through the README file or searching \textit{Google Scholar} to find research articles related to these tools.
We counted pull requests given prior work that suggests these are generally more representative of actual engagement than looking at total commits~\cite{bertoncello2020pull}. We used \textit{GraphQL} to explore the trend of pull requests evolving for last 10 years.
Similar to pull requests, we chose forking because it is a form of engagement that signals a desire to potentially build on or contribute to an existing project~\cite{zhou2019fork}.
In our work, we considered open, closed, merged pull requests.
We chose watches and stars as more social forms of engagement, where stars signal interest in or satisfaction with the project or the topic of its content (e.g., in our case machine learning fairness)~\cite{begel2013social, borges2018s} and watches signal interest in updates regarding a given project~\cite{dabbish2012social}.
We evaluated these metrics for both \textit{Original Toolset} and \textit{Emergent Toolset}.

\subsection{Evaluating Tool Maintenance (RQ2)} \label{eval_maintenance}

For both the \textit{Original Toolset} and \textit{Emergent Toolset}, we analyzed repository artifacts relevant to the maintenance of the tool. We categorized each repository with respect to its  maintenance level. For this task, we leveraged prior work from Coelho et al.~\cite{coelho2020github} that provides infrastructure for classifying open source repositories into one of three maintenance levels:

\begin{itemize}
    \item \textbf{Active:} ``Active'' refers to repositories that had at least one commit in the last 6 months relative to the date when the data was collected.
    \item \textbf{Inactive:} ``Inactive'' refers to repositories that were classified as unmaintained, following the criteria outlined in the previously published paper\cite{coelho2017modern}. This classification typically applies to repositories declared by their maintainers as no longer actively maintained and has little activity throughout the repository in the last 2 years.
    \item \textbf{Archived:} ``Archived'' indicates that the repository's owner has set the repository to a ``read-only'' state on GitHub regardless of when the last activity was.
\end{itemize}

For this study, we used a dataset from related work\cite{coelho2020github} with the labels mentioned above, which consists of 1003 repositories to classify the maintenance level of these repositories.
In their research, the authors experimented with various machine learning models and found that Random Forest yielded the most accurate classification results. 
Based on this, we selected Random Forest as our classifier to achieve the most reliable predictions. 
The features we included for training are \textit{forks}, \textit{total issues}, \textit{closed issues}, \textit{open pull requests}, \textit{closed pull requests}, \textit{merged pull requests}, \textit{total commits}, \textit{max days without a commit}, \textit{most active developer commits}, \textit{contributors}, \textit{owner projects}, and \textit{owner commits}.
We used the Github API~\footnote{\url{https://docs.github.com/en/rest?apiVersion=2022-11-28}} to fetch these features for each of the repositories. 
We scoped our mining to the last 24 months and, as with prior work, collected each feature in the interval of 3 months~\cite{coelho2020github}. Like their approach, we conducted a hierarchical clustering analysis to eliminate features with correlations exceeding 0.7. This process resulted in a substantial data reduction, in cases of higher correlation.
Our test set was comprised 61 fairness-related repositories, which we classified using the trained model. After obtaining the classification results, we proceeded with further analysis. 
For further analysis, we analyzed the correlation between the hyperparameters using Mann-Whitney U test for significance checking the \textit{p value} against 0.05. As our data is not normally distributed we used this test. We also considered the reported issues to analyze the maintenance level of the projects according to the work\cite{de2019identifying}. We used the GitHub API to calculate the mean time to resolve the issues where we considered the closed issues from each repository. We also analyzed them by default GitHub issue labels (e.g., enhancement, documentation, bug) to understand what kind of issues take longer to resolve in these projects. To gain insights into the maintainability of these fairness repositories, we analyzed them using keywords such as \textit{API Update}, \textit{API}, \textit{Endpoint}, \textit{Integration}, \textit{Backward Compatibility}, \textit{Deprecation}, \textit{Refactor}, \textit{Changelog}, and \textit{Library Update}. This approach was inspired by prior works \cite{yu2005measuring}. We used the GitHub API to mine issues and pull requests containing these keywords within the fairness repositories. We manually examined the retrieved pull requests and issues to understand how effectively these repositories managed changes over time.

\subsection{Evaluating Project Lifespan(RQ3)}
To assess project lifespan, we analyzed how long the projects survived. For this study, we reviewed each project to determine how long it has remained active since its inception.
We measured the timespan of the last commit since their initial commit to measure how long they are remained active and when was last time since developers committed to an individual repository.
This approach allowed us to gain insights into the sustainability and lifespan of various types of open source fairness tool repositories.

\section{Findings}
\label{result}

Our efforts provide valuable insights into the landscape of
and engagement with fairness tools in open source. In this
section, we discuss our findings for each of our research
questions. Our study artifacts and data are publicly available for replication and reuse.~\footnote{\url{https://anonymous.4open.science/r/Fairness-Tool-Maintenace--BEB1}}

\subsection{Open Source Fairness Tool Dataset (RQ0)}

Our efforts yielded a total of 62 open source fairness tools in Figure~\ref{industry}.
The tools we found available on GitHub, grouped by their development origin (which we discuss in more detail in Section~\ref{subsec:RQ1}). 
We denote the tools from our \textit{Original Toolset} with an asterisk (*); all others belong only to the \textit{Emergent Toolset}. Many of the fairness tools we encountered in our study, and most of the tools investigated in prior work~\cite{lee2021landscape,mim2023taxonomy}, are general purpose fairness tools that could be applicable to any domain. 
However, our findings suggest there are (probably necessary) efforts to develop fairness support specific to a given domain (e.g., healthcare). 
Furthermore, the few tools we found that provide domain-specific support came from either industry (\textit{KenSciResearch/fairMLHealth}, \textit{JohnSnowLabs/langtest}, ) or an individual (\textit{dbountouridis/siren}, \textit{kamyabnazari/fair-energy-ai}, \textit{martinetoering/Embetter}).\\

\begin{figure}[ht]
  \includegraphics[width=6cm]{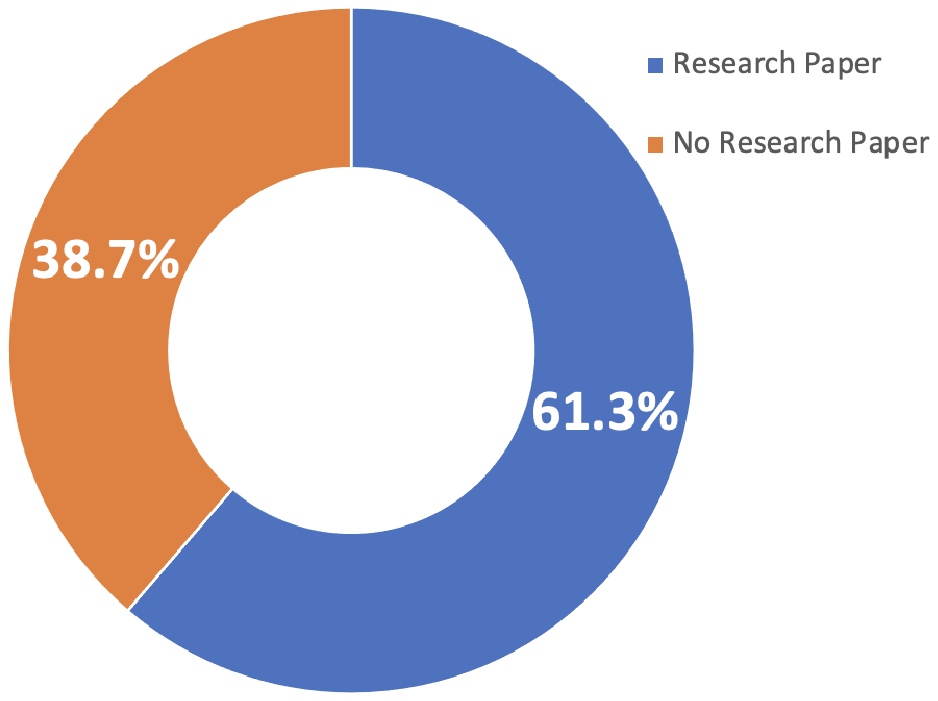}\centering
  \caption{Availability of Research Articles with the tools} 
  \label{paper}
\end{figure}

We found that 43\% of the tools originated from industry, 40\% from academy and 17\% are individual efforts.
We also analyzed each contribution to determine if there was an associated research article and found that the majority of the projects in our dataset have published an article (Figure~\ref{paper})
These findings form the foundation for our analyses on the remainder of our research questions.



\subsection{Fairness Tool Community Engagement (RQ1)}\label{subsec:RQ1}

To answer RQ1, we measured and analyzed  \textit{social engagement} through the number of stars and watches on each tool repository along with \textit{project engagement} based on the number of forks and watches.

\subsubsection{Social Engagement}

\begin{figure}[ht]
\includegraphics[width=10cm,height=9cm]{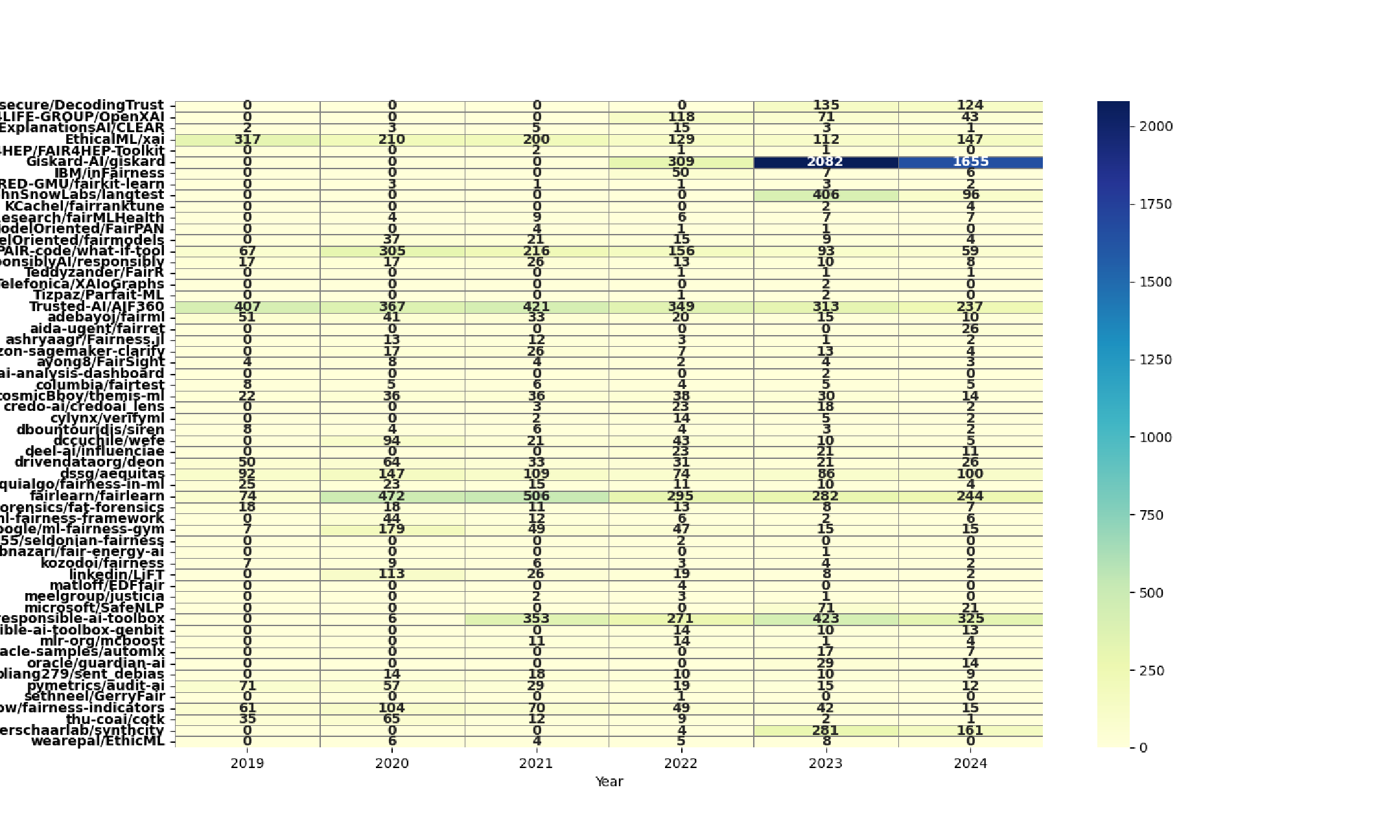}\centering
  \caption{Evolution of Project Stars (2019-2024)} 
  \label{star}
\end{figure}

To better understand social engagement with existing open source fairness tools, we first observed stars over time for each repository.
Figure~\ref{star} shows stars counts across repositories from 2019 to 2024. 
The heatmap shows trends in annual star growth, highlighting shifts in popularity and engagement over time. 
As shown, some projects, such as \textit{Trusted-AI/AIF360} and \textit{fairlearn/fairlearn}, exhibit   high star counts across this timespan, reflecting consistent community engagement, while others like \textit{PAIR-code/what-if-tool} initially spiked in popularity but have gradually declined in recent years. 
Emerging tools, including \textit{AI-secure/DecodingTrust}, \textit{vanderschaarlab/synthcity}, and \textit{Giskard-AI/giskard}, have gained significant traction recently, suggesting they may be meeting recent community needs. 
A few repositories, such as \textit{google/ml-fairness-gym}, exhibit sudden star increases during 2020-2021.
Long-standing repositories such as \textit{dssg/aequitas} and \textit{pymetrics/audit-ai} exhibit steady but moderate interest, as evidenced by their slow star growth.

The \textit{WatcherCount} distribution illustrates varied levels of social engagement, with higher counts frequently appearing in industry-backed projects, suggesting that these repositories are closely monitored for updates or releases. Examples include \textit{microsoft/responsible-ai-toolbox} (31 watchers), \textit{Trusted-AI/AIF360} (91 watchers), and \textit{dssg/aequitas} (43 watchers). These repositories, backed by renowned organizations or focusing on widely applicable solutions, usually attract a larger audience. Academic repositories like \textit{vanderschaarlab/synthcity} (13 watchers) demonstrate that research-focused tools can gather a dedicated community, emphasizing interest in innovative methodologies within the academic community.

\begin{figure}[ht]
  \includegraphics[width=10cm]{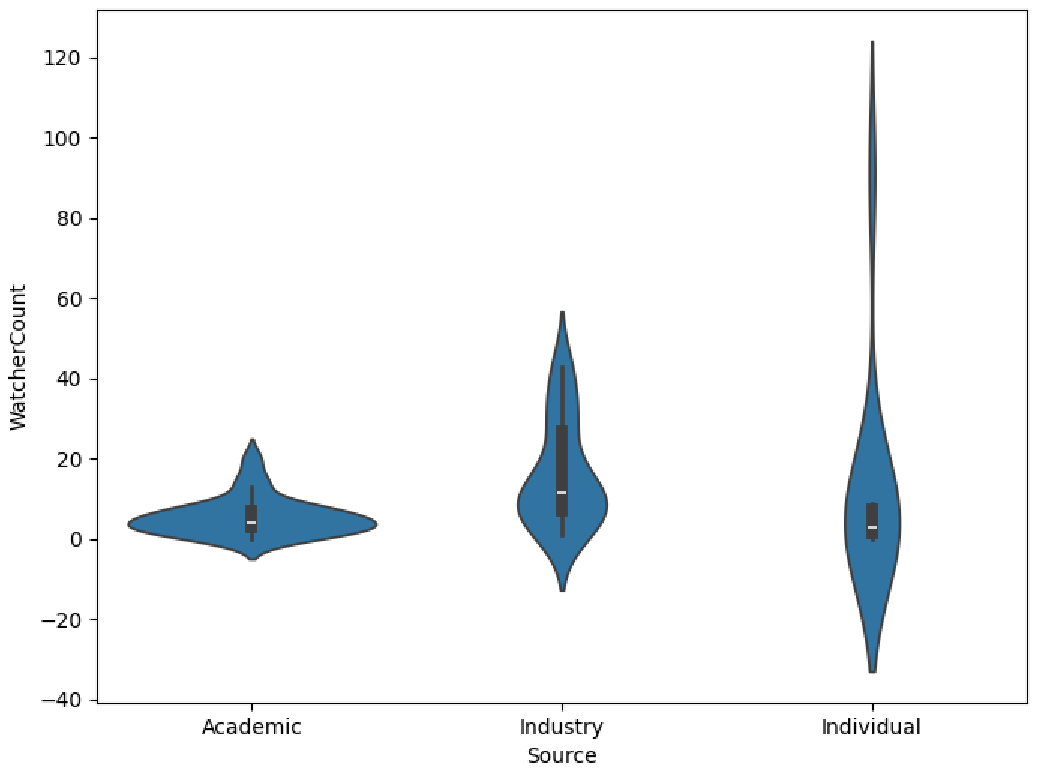}\centering
  \caption{Distribution of Watches} 
 \label{watch}
\end{figure}

\subsubsection{Project Engagement}
\begin{figure*}[ht]
  \includegraphics[width=19cm,height=12cm]{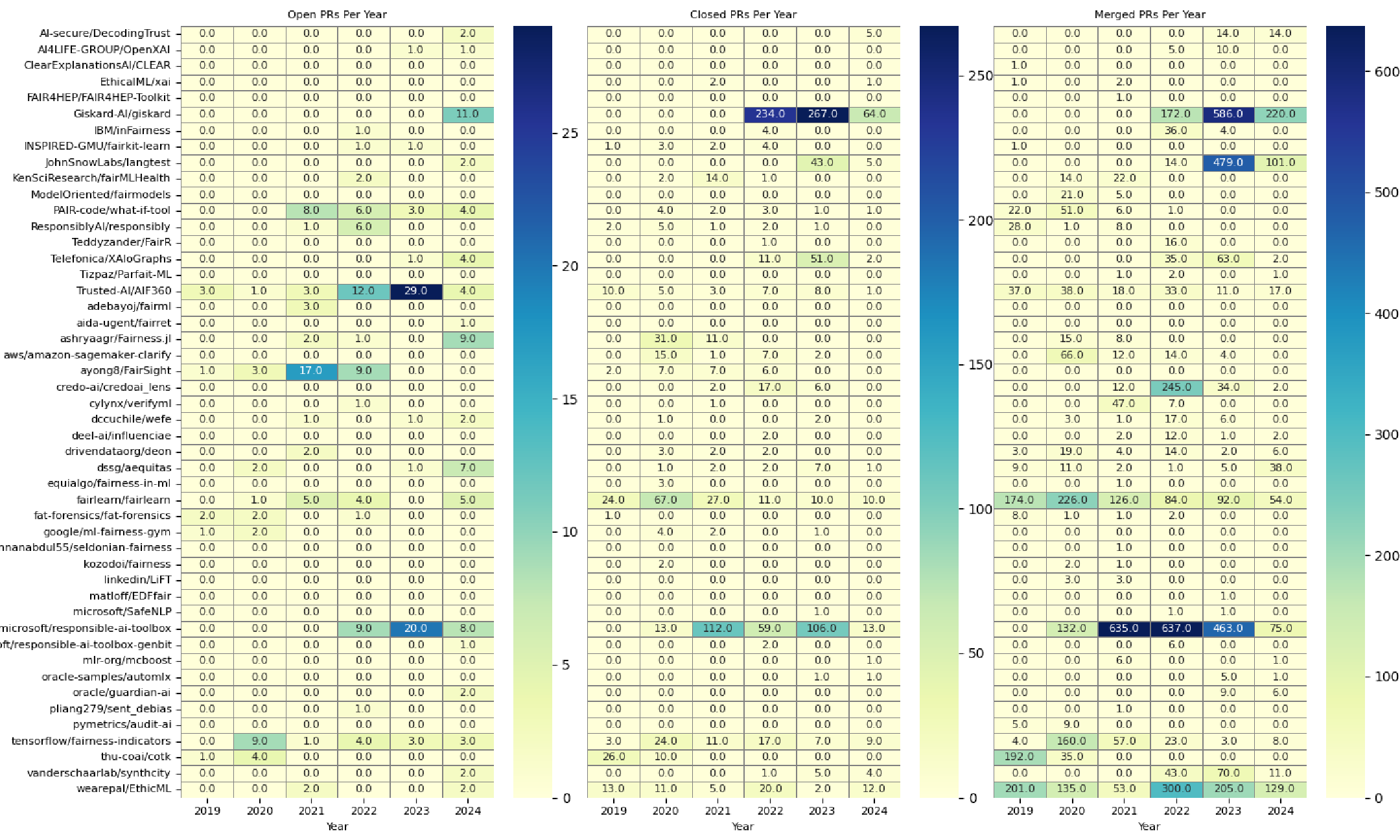}\centering
  \caption{Evolution of Open/Closed/Merged PR for the projects (2019-2024)} 
  \label{open_close_merge}
\end{figure*}

To characterize deeper engagement with fairness tool projects, we analyzed the annual counts of open, closed, and merged pull requests (PRs) to observe distinct patterns of development and maintenance activity (Figure~\ref{open_close_merge}).
While we initially collected data to analyze trends from 2016-2024, many repositories had little to no pull request activity prior to 2019.
Overall, we found the highest pull request activity across fairness tools with origins in industry.
For instance, repositories such as \textit{microsoft/responsible-ai-toolbox} demonstrate robust annual PR activity, highlighted by substantial numbers of closed and merged PRs, with a peak of 112 closed PRs in 2021 and an impressive 637 merged PRs in 2022, indicating strong community engagement and consistent maintenance.
Even with older repositories like \textit{dssg/aequitas} and \textit{fairlearn/fairlearn}, we found stable PR activity each year, reflecting sustained community engagement. 
Projects such as \textit{tensorflow/fairness-indicators} and \textit{PAIR-code/what-if-tool} stand out for their high rates of merged PRs relative to opened PRs, indicative of engagement from both ends (those making contributions and those that maintain the tool). 
In contrast, we found that most of the projects that came from academia or an individual, such as \textit{INSPIRED-GMU/fairkit-learn} and \textit{kozodoi/fairness}, exhibit lower, sporadic PR activity, potentially indicating niche interest or less frequent updates.

Some repositories, like \textit{Giskard-AI/giskard} and \textit{vanderschaarlab/synthcity}, display recent spikes in PR contributions, suggesting increased visibility or significant updates that have higher values for social engagement metrics. For instance, the repository \textit{Giskard-AI/giskard} has the greatest number of stars among all the fairness-related repositories.



\begin{figure}[ht]
  \includegraphics[width=10cm]{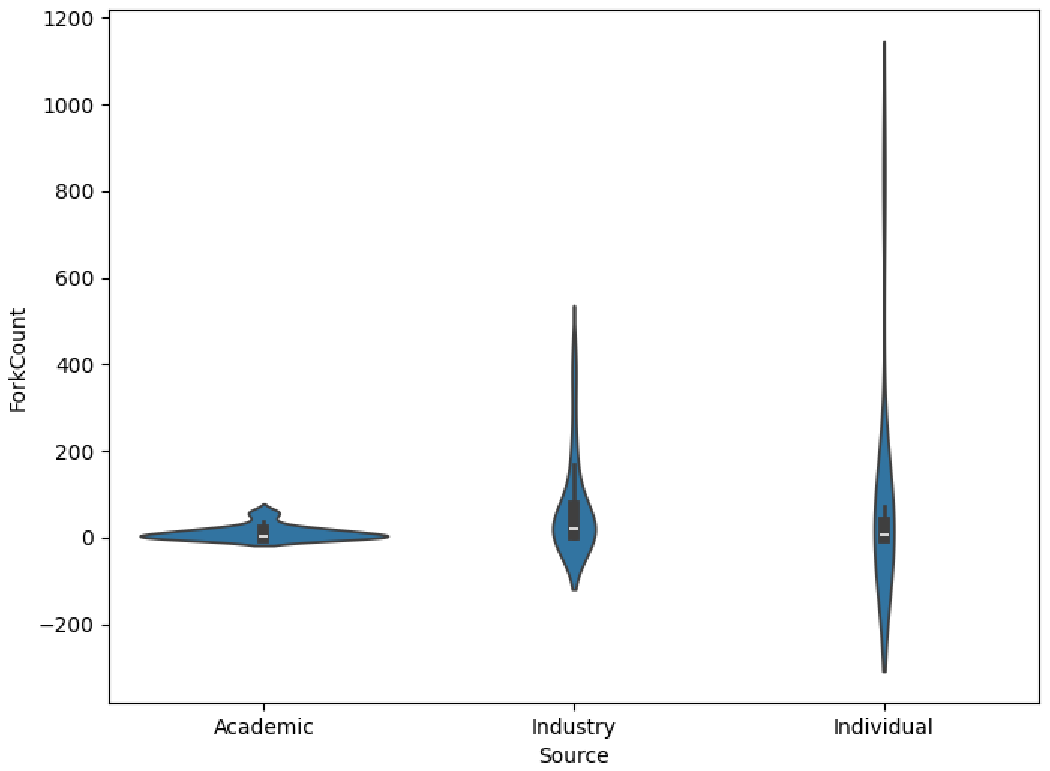}\centering
  \caption{Distribution of  Forks}  
  \label{fork}
\end{figure}

Similar to pull requests, as shown in Figure~\ref{fork}, we found that industry projects generally exhibit higher fork counts, as observed with notable repositories like \textit{microsoft/responsible-ai-toolbox} (361 forks), \textit{Trusted-AI/AIF360} (838 forks), and \textit{fairlearn/fairlearn} (418 forks). 
Academic and individual repositories, such as \textit{ayong8/FairSight} and \textit{KCachel/fairranktune}, generally have lower fork counts, reflecting a smaller community or more specialized interests. 
However, some academic repositories like \textit{AI-secure/DecodingTrust} (55 forks) and individual ones like \textit{adebayoj/fairml} (74 forks) also demonstrate considerable community engagement, showing that impactful or popular repositories can attract significant interest across all source.


\subsection{Fairness Tool Maintenance (RQ2)}

To acquire insights into how actively open source fairness projects are being maintained, we leveraged prior works to classify each project based on its activity. 
We found that only 32.2\% of the projects in our dataset would be considered active.
The majority of the remainder would be considered inactive (62.7\%) with a handful set to read-only (archived) status (5.1\%).
For the \textbf{Active} projects, we found high volume of \textit{total issues}, \textit{closed issues}, and \textit{closed PRs}.
We also found that \textbf{Active} repositories exhibit more balance across \textit{open PRs}, \textit{closedPRs}, and \textit{merged PRs}.
For \textbf{Inactive} projects, we observed fewer contributions (lower \textit{owner commits}, \textit{contributors}, and \textit{most active dev commits}).
We also observed longer periods of inactivity (\textit{max days without commit}) and an unusual correlation between the number of \textit{owner commits} and the maintenance status where we found that the mean value of \textbf{owner commits} is higher for \textbf{Inactive} projects compared to active ones.

We used the Mann-Whitney U test to determine whether statistically significant differences exist between parameters of projects classified as \textit{Active} (20) and \textit{Inactive} (36). 
As shown in Table~\ref{table_mann_whitney}, there are significant differences with respect to \textit{total issues}, \textit{open PRs}, \textit{closed PRs}, \textit{merged PRs}, \textit{total commits}, \textit{most active dev commits}, and \textit{contributors} between active and inactive projects with p-value not more than 0.05 marked with(*). The Mean Difference column represents the difference in mean values between the Active and Inactive groups for each row. Negative values in metrics like Max Days Without Commit and Owner Commits indicate that the mean values are higher for Inactive groups. Interestingly, Inactive repositories exhibit a higher number of owner commits.



\begin{figure}[ht]
  \includegraphics[width=9cm]{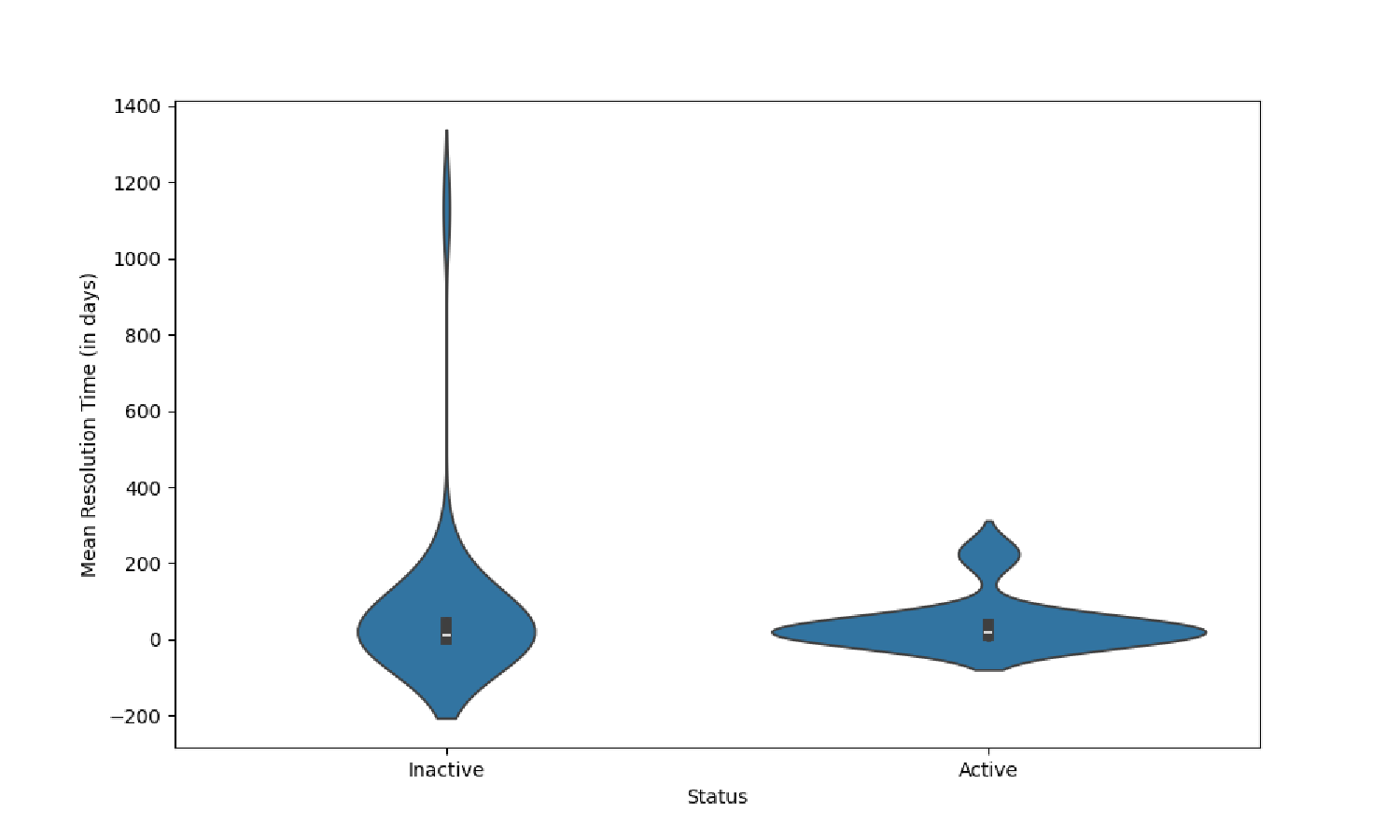}
  \caption{Distribution of Mean Issue Resolution Time} 
  \label{mean_issue}
\end{figure}

\begin{table}[ht]
\centering
\caption{Mann-Whitney U Test for Differenting Active and Inactive Projects}
\label{table_mann_whitney}
\begin{tabular}{lrlr}
\toprule
\textbf{Parameter} &  \textbf{U Statistic} & \textbf{p-value}  & \textbf{Mean Difference} \\
\midrule

Forks  & 462.0 & 0.0819 & 91.266667 \\

\rowcolor{Gray}
\textbf{Total Issues}  & \textbf{520.5} & \textbf{0.0062*} & \textbf{321.15}\\

\textbf{Closed Issues}  & \textbf{526.5} & \textbf{0.0044*} & \textbf{287.1667}\\

\rowcolor{Gray}
\textbf{Open PRs}  & \textbf{484.5} & \textbf{0.0241*} & \textbf{6.37}\\

\textbf{Closed PRs}  & \textbf{519.5} & \textbf{0.0062*} & \textbf{225.99} \\

\rowcolor{Gray}
\textbf{Merged PRs}  & \textbf{535.5} & \textbf{0.0025*} & \textbf{195.27}\\

Total \textbf{Commits}  & \textbf{536.5} & \textbf{0.0026*} & \textbf{512.29} \\

\rowcolor{Gray}
Max Days Without Commit  & 331.0 & 0.6260 & -81.38 \\

\textbf{Most Active Dev Commits}  & \textbf{487.0} & \textbf{0.0305*}  & \textbf{113.45}\\

\rowcolor{Gray}
\textbf{Contributors}  & \textbf{563.5} & \textbf{0.0005*} & \textbf{13.74}\\

Owner Projects  & 382.5 & 0.4834 & 0.30 \\

\rowcolor{Gray}
Owner Commits  & 269.5 & 0.0586 & -20.24 \\

\bottomrule
\end{tabular}
\end{table}
We also analyzed and compared the mean issue resolution time for active and inactive projects. 
As shown in Figure~\ref{mean_issue}, \textbf{Active} projects generally exhibit moderate resolutions times (often under 50 days), with a few exceptions reaching over 200 days (e.g., \textit{microsoft/responsible-ai-toolbox} and \textit{dssg/aequitas}). 
\textbf{Inactive} projects displayed a wider range of resolution times, from very low values (0 days) indicating little activity or quick closures, to extremely high values, such as 1131 days for \textit{pliang279/sent\_debias}. 
The wide variation in resolution times further emphasizes, and possibly explain, the differences in engagement that indicate active or inactive status.


To better understand the kinds of issues resolved quickest (and slowest), we analyzed issue labels across repositories. 
Of the 61 projects in our dataset, only 19 labeled their issues (11 \textbf{Active}, 6 \textbf{Inactive}, 2 \textbf{archived}). 
Upon further investigation, we found only 3 repositories labeled all of their issues, 2 of which are active: \textit{Langtest}, \textit{Responsible-ai-toolbox} and \textit{cotk}.
Across repositories, we found a range of different labels both built-in and created.
To ensure we properly interpret the labelings, we focused our analysis on the default labels provided by GitHub.

For the \textbf{Active} projects, we found that same label can take both higher and lower than average issue resolution time (e.g., \textit{deel-ai/influencia, fairlearn/fairlearn}).
'Enhancement' issues took the longest for most of the repositories (except \textit{AI4LIFE-GROUP/OpenXAI}) and appeared most frequently as a label among the fairness tool repositories.
`Documentation' issues are also typically resolved faster within the projects with the exception of \textit{drivendataorg/deon}. 
Issues labeled 'bug' have mixed resolution times, being higher for some (\textit{e.g, vanderschaarlab/synthcity}, \textit{AI-secure/DecodingTrust}) and lower for others (\textit{e.g, Trusted-AI/AIF360, JohnSnowLabs/langtest}). 
We also found mixed resolution times for issues labeled `good first issue,' which signals to potential newcomers there is a lower barrier to entry. 
For the \textbf{Inactive} projects, issues labeled 'enhancement', `documentation', `question', and `good first issue' took the  longest for most of the repositories.
We found that `bug' issues were resolved relatively quickly, with the exception of \textit{ModelOriented/fairmodels}. \\

We also found some fairness-oriented keywords in the labels as well. Which are \textbf{ethics-checklist} found in \textit{drivendataorg/deon} taking less than mean issue resolution time and \textbf{fairness} in \textit{microsoft/responsible-ai-toolbox} taking longer than mean issue resolution time of their own repository.

\noindent\textit{Efforts Toward Compatibility and Long-Term Maintability.}
Our analysis open source fairness repositories revealed a strong emphasis on maintainability.
We found high frequencies of API-related terms across repositories (including 613 mentions of API updates) signaling efforts towards software robustness and adaptability, such as \textit{KenSciResearch/fairMLHealth} adding regression features and integration enhancements. 
There were also numerous mentions of refactoring (251 occurrences) for improving code clarity and efficiency, exemplified by \textit{INSPIRED-GMU/fairkit-learn}, which resolved validation bugs and streamlined code. Repositories also exhibited significant integration activities (367 mentions) with modern frameworks, such as \textit{google/ml-fairness-gym} transitioning to TensorFlow 2.5.x to maintain compatibility. Additionally, library update references (285 occurrences) indicate attention to security risks, as demonstrated by \textit{oracle/guardian-ai}, which upgraded dependencies like \textit{scipy} and \textit{scikit-learn}.

We found 38 occurrences of the keyword 'backward compatibility' across all repositories, such as in \textit{dccuchile/wefe}, which addressed missing WEAT words and updated torch for compatibility. Similarly, there were 129 occurrences of the keyword 'deprecation' across all repositories, with \textit{wearepal/EthicML} focusing on dependency upgrades like gitpython and requests to prevent deprecated issues.\\

We also observed a high frequency of changelog references (543), particularly in \textit{ayong8/FairSight} and \textit{Trusted-AI/AIF360}.
While addressing the deprecation of outdated features ensures code efficiency, mentions of endpoint modifications (92 occurrences), such as \textit{pymetrics/auditai}'s updates to fairness metric evaluations, indicate ongoing API refinement for increased usability and relevance. 
 Comprehensive documentation improvements, such as those we found in \textit{PAIR-code/what-if-tool}, support long-term stability and collaboration, reflecting a consistent effort to make these tools more accessible and maintainable.

\subsection{Project Lifespan (RQ3)}
To evaluate the lifespan of a project that is how many years it remains active, we observed the number of years the project has been active based on \textit{active commits} and the interval since the last commit occurred.

\begin{figure*}[ht]
  \centering
  \includegraphics[width=16cm]{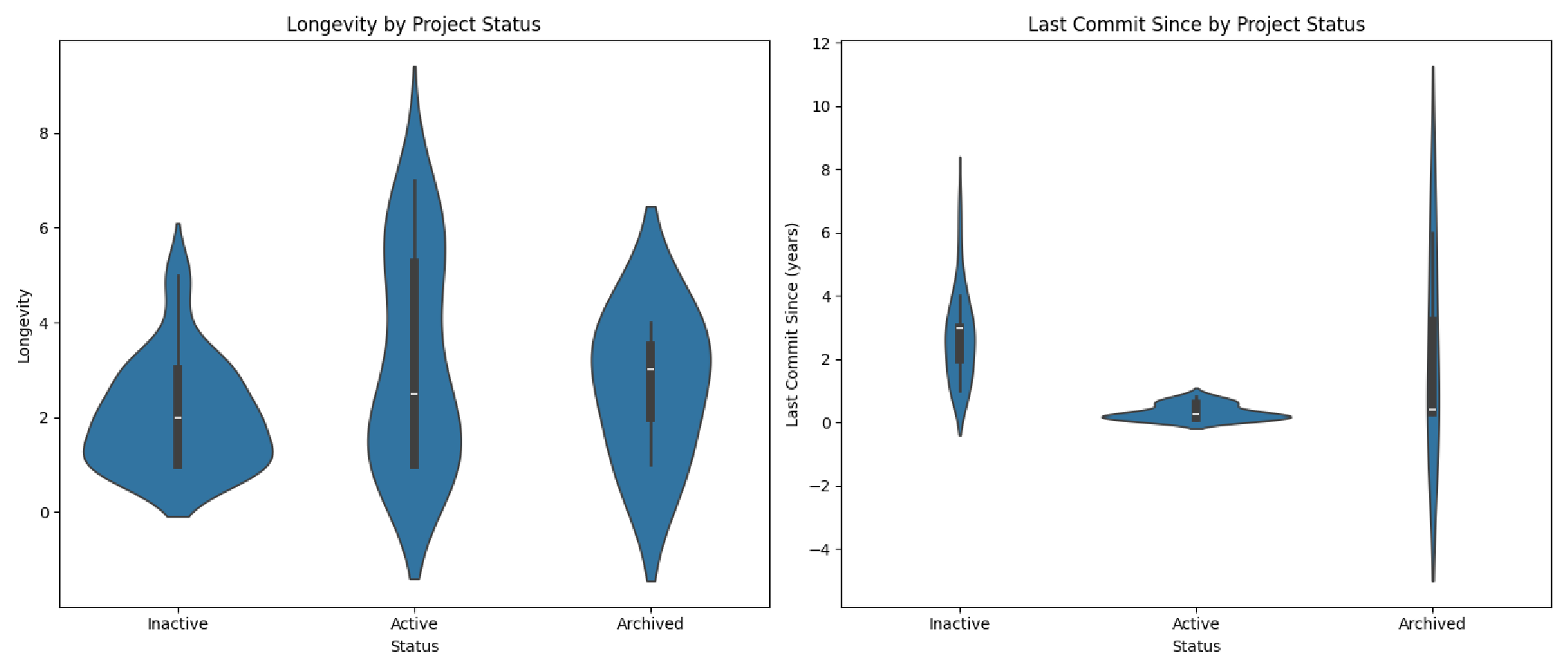}
  \caption{Distribution of Lifespan of the Projects}
  \label{longevity}
\end{figure*}

As shown in Figure~\ref{longevity}, \textbf{Active} projects generally exhibit recent commit activity, with \textit{Last Commit Since} values frequently below one year, suggesting ongoing maintenance.
Many \textbf{Active} projects also have a lower project lifespan (1-3 years), indicating they are relatively new but consistently updated. 
In contrast, \textbf{Inactive} repositories tend to have longer periods since their last commit, often 2-4 years, reflecting reduced or paused development. 
Some \textbf{Inactive} projects, despite higher lifespan, had limited recent contributions. 
\textbf{Archived} repositories, while fewer, typically exhibited  high \textit{Last Commit Since} values, underscoring their discontinued status. 
Our findings suggest distinct patterns of activity and maintenance based on repository status, with \textbf{Active} projects demonstrating more recent updates and shorter gaps in commits than \textbf{Inactive} or \textbf{archived} ones. 
The \textbf{archived} repositories exhibit varying levels of lifespan, generally indicating that they have been in existence for a moderate to long time (ranging from 3 to 4 years). 

\section{Discussion}
\label{discussion}
Our findings build on prior efforts aimed at characterizing the landscape of fairness tools and the ways in which we can encourage and support integration in practice. In this section, we discuss insights for research and practice regarding open source fairness tools .

\subsection{Increasing Visibility of Fairness Interventions}

The goal of our work is support the sustainability of open source fairness tool projects by evaluating their maintenance and community engagement history. 
While prior work suggests tool maintenance may be a relevant consideration with respect to interest in or engagement with a given tool~\cite{johnson2023make}, our findings indicate that in open source consistent maintenance may not be enough to attract and engage potential users.
Maturity may be a relevant factor, as we found in some cases that the well maintained repositories with high engagement were also those that had been active longer. 
Furthermore, newer projects may lack the visibility of older projects. 
Future research can work to solidify the factors outside of high engagement that may signal a sustainable project. 



Another known issue impacting engagement with fairness tools is lack of awareness~\cite{lee2021landscape, mim2023taxonomy}.
Aside from any implications regarding engagement and awareness, we hope this work can serve as an opportunity to raise awareness  for researchers and practitioners regarding the vast landscape of open source fairness tools that can be investigated and integrated. 
As emphasized by insights from prior work, peers and communities play an important role in tool awareness and trust~\cite{murphy2011peer, johnson2023make}.  
All these efforts highlight a need to support increased awareness regarding the variety of fairness tools available; it may be that some of the lesser known or less visible tools provide useful or supplementary support compared to those that are more familiar.

\subsection{Supporting Domain-Specific Fairness Considerations}

Many of the fairness tools we encountered in our study, and most of the tools investigated in prior work~\cite{lee2021landscape,mim2023taxonomy}, are general purpose fairness tools that could be applicable to any domain. 
In fact, repositories like \textit{AI Fairness 360} include examples and datasets across different domains (e.g, banking~\footnote{\url{https://github.com/Trusted-AI/AIF360/blob/main/aif360/datasets/bank_dataset.py}} and education~\footnote{\url{https://github.com/Trusted-AI/AIF360/blob/main/aif360/datasets/law_school_gpa_dataset.py}}).
However, our findings suggest there are (probably necessary) efforts to develop fairness support specific to a given domain (e.g., healthcare). 
 For instance, in our dataset of tools we found \textit{KenSciResearch/fairMLHealth} which is geared towards healthcare, \textit{JohnSnowLabs/langtest} which focuses on language model testing, \textit{dbountouridis/siren} which  addresses fairness in recommender systems for online news, \textit{kamyabnazari/fair-energy-ai} which is oriented towards energy management, and \textit{martinetoering/Embetter} which centers on word embeddings. 
This suggests that in practice there may be gaps with respect to fairness tooling that can support the unique considerations that come with leveraging machine learning solutions in different domains.
We also observed that the few tools we found that provide domain-specific support came from either industry or an individual. 
This further supports the need for more domain-specific fairness support and provides an opportunity for research to better align with practice by exploring domain-specific solutions.

\subsection{Fostering Fairness Tool Stability in Open Source}

While we found mixed engagement and maintenance trends across the projects we analyzed, the projects that came from industry or non-academic organizations stood out as the most stable. 
The most obvious explanation is the visibility of the organizations behind these tools, especially given prior work suggesting reputation plays a role in the trust engineers have in adoption and use of AI or ML as a solution~\cite{johnson2023make}.
However, given the examples of stable academia- and individually-backed projects in our dataset, there is likely more to stability in providing open source fairness support than brand reputation.

Trends in our data suggest that providing niche, high demand, or domain-specific solutions may be another way to engage potential users and contributors in open source fairness support. 
For example, industry projects like \textit{JohnSnowLabs/langtest} and \textit{microsoft/SafeNLP} experienced massive surges in engagement and maintenance activities between 2023 and 2024, which coincides with the rise of large language models (LLMs) and release of technologies ChatGPT~\cite{xi2023rise}. 
We also found domain-specific academic (\textit{ayong8/Fairsight}) and individual (\textit{dbountouridis/siren}) projects that maintained low, but consistent engagement and maintenance activity between 2019 and 2024. 
This suggests that some may be seeking niche solutions rather than all purpose fairness tooling, emphasizing that value and perceived relevance may play a role in stability as well. 
Future research on fairness tools in open source can help answer the question regarding other open source fairness project factors that contribute to stable engagement and the ability or need to actively engage in maintenance activities.

\section{Threats to validity}
\label{validity}
We designed our study to explore the landscape of machine learning fairness support available in the open source. 
While we took every precaution to design a sound experiment, there are some potential threats to the validity of our insights.

\subsection{External Validity}
Our research focused on open source fairness projects made publicly available on GitHub. Therefore, our insights may not scale beyond open source fairness repositories. 
Furthermore, the tools in our sample is likely only a subset of tooling available to support fairness concerns. Which means our insights may not fully represent the landscape of fairness support.

\subsection{Internal \& Construct Validity}
To find the tools for our study, we leveraged keywords from the \textit{Original Toolset} documentation. 
Therefore, we may have missed potentially relevant keywords or tools we could have found with them. 
To measure engagement and maintenance, we focused on specific features of each repository. 
While we attempted to be exhaustive by leveraging prior work, there may be other factors we did not consider that may have meaningful relationships with engagement and maintenance of open source fairness tools.

\section{Conclusion}
\label{conclusion}
\label{conclude}
In this paper, we presented an analysis of 61 publicly available fairness tools in the open source community. 
Our analysis yielded novel and valuable insights into the landscape of available fairness support with respect to engagement with the broader open source community, maintenance, and stability. 
Our findings have implications for increasing and sustaining engagement with open source fairness tools and opportunities for advancing research and development in fairness tools support.


\nocite{*}

\bibliographystyle{plain}
\bibliography{bibliography}

\end{document}